\documentclass[11pt, a4paper, logo]{googledeepmind}

\usepackage[T1]{fontenc} 

\usepackage[
    natbib=true,
    backend=biber,
    style=numeric, 
    sorting=none 
]{biblatex}
\AtEveryBibitem{\clearfield{month}}
\AtEveryBibitem{\clearfield{day}}
\usepackage{csquotes}

\title{IntFold: A Controllable Foundation Model for General and Specialized Biomolecular Structure Prediction}

\correspondingauthor{
\hspace{-1.5mm} send correspondence regarding this report to contact@intfold.com}

\renewcommand{\today}{2025-07-02}

\author[1]{The IntFold Team}

\affil[1]{IntelliGen AI}

\usepackage{pdflscape}
\usepackage{xcolor}
\usepackage{textcomp}
\usepackage{rotating}
\usepackage{setspace}
\usepackage{microtype} 

\usepackage{graphicx}
\usepackage{subcaption}
\usepackage{caption}
\usepackage{longtable}
\usepackage{booktabs}
\usepackage{makecell}

\usepackage{array}
\usepackage{threeparttable}
\usepackage{multirow}

\usepackage{amsmath}
\usepackage{siunitx}

\usepackage{enumitem}
\usepackage{float}
\usepackage{seqsplit}

\usepackage{hyperref}
\hypersetup{
    colorlinks=true,
    linkcolor=blue, 
    citecolor=blue,  
    filecolor=black,
    urlcolor=blue    
}

\usepackage{tocloft}

\usepackage{etoolbox}
\makeatletter
\patchcmd{\@tocline}
    {\hfil}
    {\leaders\hbox{\hfil}\hfil}
    {}{}
\makeatother

\AtBeginDocument{

}

\begin{abstract}
We introduce IntFold, a controllable foundation model for general and specialized biomolecular structure prediction. Utilizing a high-performance custom attention kernel, IntFold achieves accuracy comparable to the state-of-the-art AlphaFold 3 on a comprehensive benchmark of diverse biomolecular structures, while also significantly outperforming other leading all-atom prediction approaches. The model's key innovation is its controllability, enabling downstream applications critical for drug screening and design. Through specialized adapters, it can be precisely guided to predict complex allosteric states, apply user-defined structural constraints, and estimate binding affinity. Furthermore, we present a training-free, similarity-based method for ranking predictions that improves success rates in a model-agnostic manner. This report details these advancements and shares insights from the training and development of this large-scale model.
\end{abstract}

\begin{document}
\sloppy 
\maketitle
\begin{refsection} 
\section*{Experimental highlights}
\begin{itemize}
    \item \textbf{State-of-the-Art Accuracy:} On the comprehensive FoldBench benchmark, IntFold's performance is comparable to AlphaFold 3 and significantly exceeds other contemporary models (Boltz-1, 2, Chai-1, HelixFold 3 and Protenix) across a diverse range of biomolecular interaction tasks.
    
    \item \textbf{Versatile Model Adaptation with Simple Tuning:} We control the model's predictions towards tasks relevant to drug screening and design by inserting small, trainable adapters while the large base model remains frozen. This efficient method allows IntFold to solve challenges where general models fail:
        \begin{itemize}
        \item Allosteric States: Fine-tuning successfully enables the prediction of specific, functionally critical allosteric conformations, a known difficulty for general-purpose models.
        \item Pocket and Epitope Guidance: The model can incorporate structural constraints for known binding pockets or epitopes, leading to more accurate predictions of protein-ligand and antibody-antigen interactions.
        \item High-Fidelity Binding Affinity Prediction: After fine-tuning the affinity data, the model accurately predicts binding affinity, demonstrating performance that surpasses Boltz-2 and significantly enhances virtual screening capabilities.
        \end{itemize}
        
     \item \textbf{Superior Attention Kernel Performance:} Development and implementation of a custom FlashAttentionPairBias kernel that outperforms standard industry kernels. It runs faster or with lower memory usage than those developed by DeepSpeed and NVIDIA.
     
    \item \textbf{Novel Model-Agnostic Ranking Method:} We developed a model-agnostic ranking method based on structural similarity. By selecting the most self-consistent structure from a diverse set of predictions, this training-free approach consistently improves the success rate over random selection.
    
    \item \textbf{Deep Insights into Large-Scale Model Behavior:} Building this model uncovered critical challenges in large-scale training. It includes data processing complications, model parametrization choices, and the origins of internal model instabilities, such as large activation magnitudes.
\end{itemize}

\clearpage

\tableofcontents
\enlargethispage{1cm} 
\thispagestyle{empty}

\section{Introduction}
\label{sec:introduction}

The prediction of three-dimensional biomolecular structures is a fundamental problem in biology and medicine. Accurate structure modeling is essential for understanding biological mechanisms, disease pathways, and accelerating therapeutic designs. This field was revolutionized by DeepMind's AlphaFold 2~\citep{jumper2021highly}, a deep learning model with unprecedented accuracy in protein monomer structure prediction. Building on this progress, AlphaFold 3~\citep{Abramson2024} was introduced in 2024 with a unified framework to predict interactions between proteins, nucleic acids, small molecules, ions, and modified residues, establishing a new standard for modeling general biomolecular assemblies. While this new architecture inspired a wave of research, independent reproductions have so far shown a significant performance gap compared to the original AlphaFold 3, highlighting the difficulty of successfully training such a model. 

While these advances in general-purpose prediction are transformative, significant challenges remain to efficiently adapt such large models for specialized applications. In fields like therapeutic design, researchers often need to test specific biological hypotheses or incorporate prior knowledge, such as the location of a known binding pocket or antibody epitopes. Furthermore, accurately capturing distinct functional states, such as the subtle yet critical conformations of allosteric proteins, remains a difficult task for general models that are not designed for such fine-grained control~\citep{olanders2024challenge}. This creates a critical need for a new class of models that combine state-of-the-art accuracy with a high degree of user-driven controllability.

To address these challenges, we introduce IntFold\footnote{The IntFold server and source code are available at \url{https://server.intfold.com/} and \url{https://github.com/IntelliGen-AI/IntFold}, respectively.}, a controllable foundation model designed for both high-accuracy general prediction and specialized, user-guided tasks. IntFold achieves two primary goals: first, it provides predictive accuracy for a wide range of biological interactions comparable to the state-of-the-art; second, it offers exceptional user control for specialized applications critical to drug screening and design. This is accomplished by inserting lightweight, trainable ``adapter'' modules while keeping the large base model frozen. 

In this report, we detail IntFold's key contributions. We first demonstrate its performance against leading methods on a comprehensive benchmark. We then showcase its versatility in several applications critical for drug design, such as modeling allosteric states and predicting binding affinity. We also introduce two technical innovations: a custom attention kernel that is more efficient than standard industry implementations and a novel confidence score designed to provide a more fine-grained and reliable quality assessment for challenging targets like antibody-antigen complexes. Finally, we share practical insights gained during development with respect to data curation, parametrization, and the root causes of training instability.

\begin{figure}[t]
    \centering
    \includegraphics[width=\linewidth]{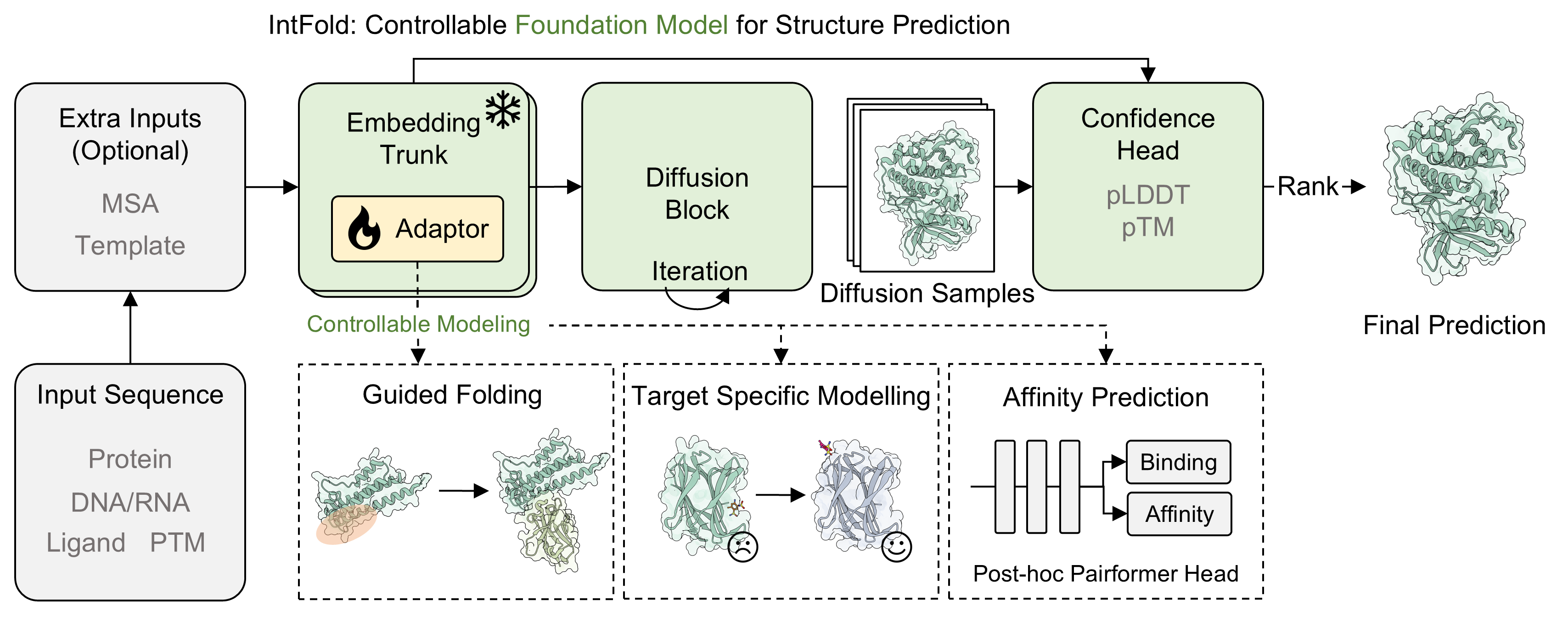}
    \caption{Diagram of the model's architecture. The model processes input sequences and optional MSAs/templates through an Embedding Trunk. Controllable modeling is achieved by inserting a modular Adaptor into this trunk. A Diffusion Block then iteratively generates structure samples, which are ranked by a Confidence Head (pLDDT, pTM) to yield the final prediction. The bottom panel shows how adapters enable specialized downstream tasks like Guided Folding, Target-Specific Modeling, and Affinity Prediction.}
    \label{fig:main_figure}
\end{figure}

\section{Benchmarking}

To comprehensively evaluate the performance of IntFold, we conducted a rigorous evaluation on FoldBench~\citep{xu2025foldbench}. We compared IntFold against several leading methods, including Boltz-1,2~\citep{wohlwend2024boltz, passaro2025boltz}, Chai-1~\citep{chai2024chai}, HelixFold 3~\citep{liu2024technical}, Protenix~\citep{bytedance2025protenix} and AlphaFold 3~\citep{Abramson2024}.\footnote{Due to license limitations, performance metrics for AlphaFold 3 are cited directly from the original benchmark. For other models, results are from FoldBench if available; otherwise, we reran them ourselves.} The following sections detail the performance across different biomolecular categories.

\quad
\\
\quad
\\
\quad
\\
\quad

\begin{figure}[H]
\centering
\includegraphics[width=\linewidth]{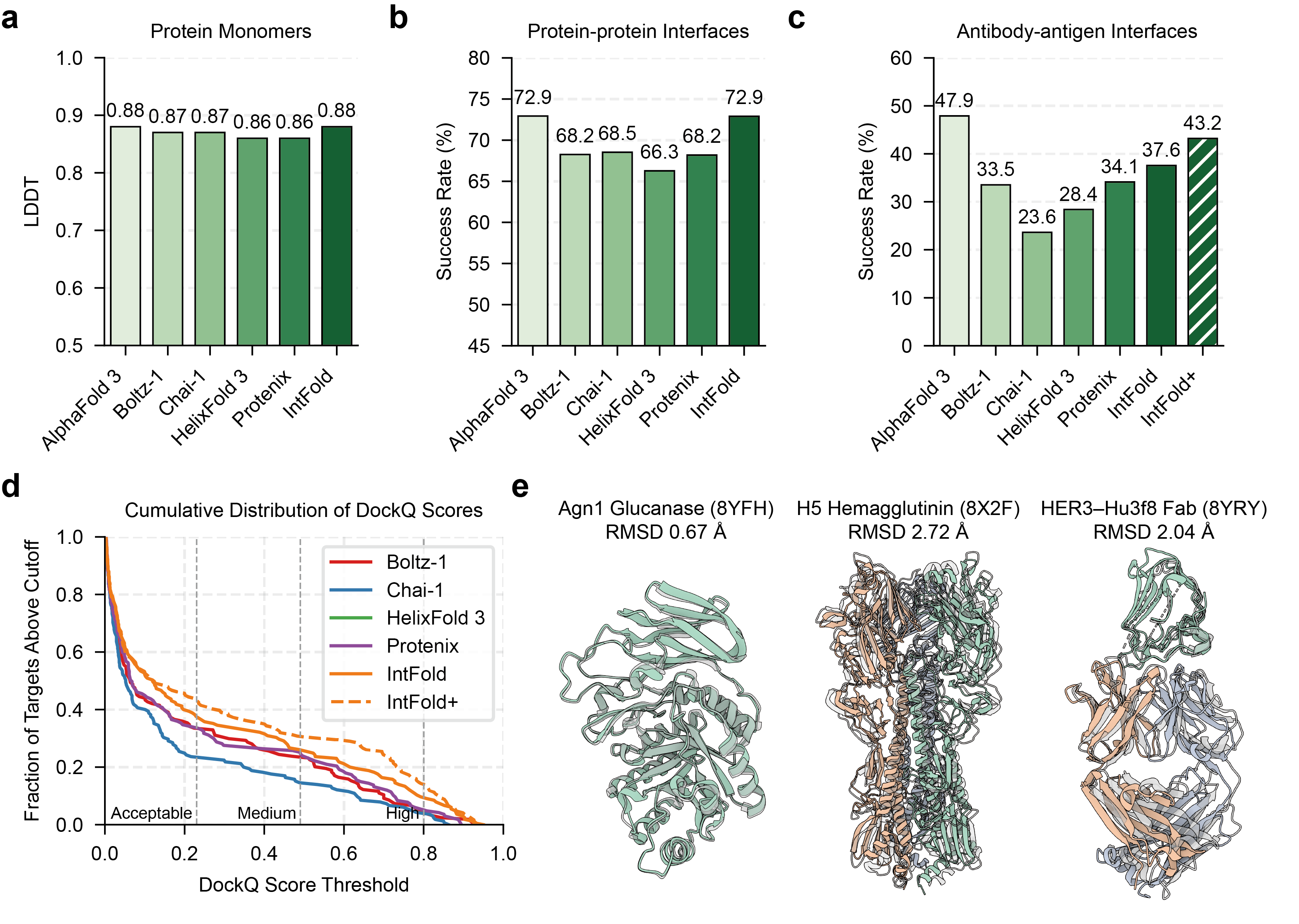} 
\caption{(a) Mean LDDT scores for protein monomer prediction on FoldBench. (b) Success rates for protein-protein interface prediction. (c) Success rates for antibody-antigen interface prediction. (d) Cumulative distribution of DockQ scores for antibody-antigen complexes, showing the fraction of targets predicted above a given quality threshold. (e) Predictions of newly released protein targets: (left) a previously unsolved yeast enzyme Agn1 Glucanase (8YFH); (center) an H5N1 hemagglutinin complex (8X2F); (right) a novel HER3-targeting antibody-drug conjugate (8YRY).}
\label{fig:protein_bench}
\end{figure}


\subsection{Proteins and Antibody-Antigen Complexes}

We first evaluated the model on protein-related systems. For protein monomers, IntFold achieves a mean LDDT score of 0.88, matching the performance of AlphaFold 3 and remaining highly competitive with the other tested methods, demonstrating strong single-chain folding ability (Figure~\ref{fig:protein_bench}a). For protein-protein interactions, IntFold reaches a success rate of 72.9\%, again matching AlphaFold 3 (72.9\%) and significantly outperforming the next best contemporary method, Chai-1 (68.5\%) (Figure~\ref{fig:protein_bench}b).

A key focus was on antibody-antigen (Ab-Ag) complexes, a modality critical for immunology and therapeutics where previous models have often struggled. A significant performance gap exists between AlphaFold 3 and other methods for this task. As shown in Figure~\ref{fig:protein_bench}c, the general IntFold model shows a success rate of 37.6\%, closing the gap to AlphaFold 3 (47.9\%). Furthermore, an enhanced model, IntFold+, improves the success rate to 43.2\%, reaching performance comparable to AlphaFold 3. Moreover, the cumulative distribution of DockQ scores shows that IntFold's advantage is consistent across all quality thresholds, producing more high-quality predictions than other methods at every threshold (Figure~\ref{fig:protein_bench}d).

To demonstrate its predictive power on prospective challenges, we validated IntFold on several complex targets newly released in 2025. As illustrated in Figure~\ref{fig:protein_bench}e, the model successfully predicted the structures of a previously unsolved yeast enzyme (8YFH), an H5N1 hemagglutinin complex (8X2F), and a novel HER3-targeting antibody (8YRY).

\subsection{Protein-Ligand Interactions}
\begin{figure}[H]
\centering
\includegraphics[width=\linewidth]{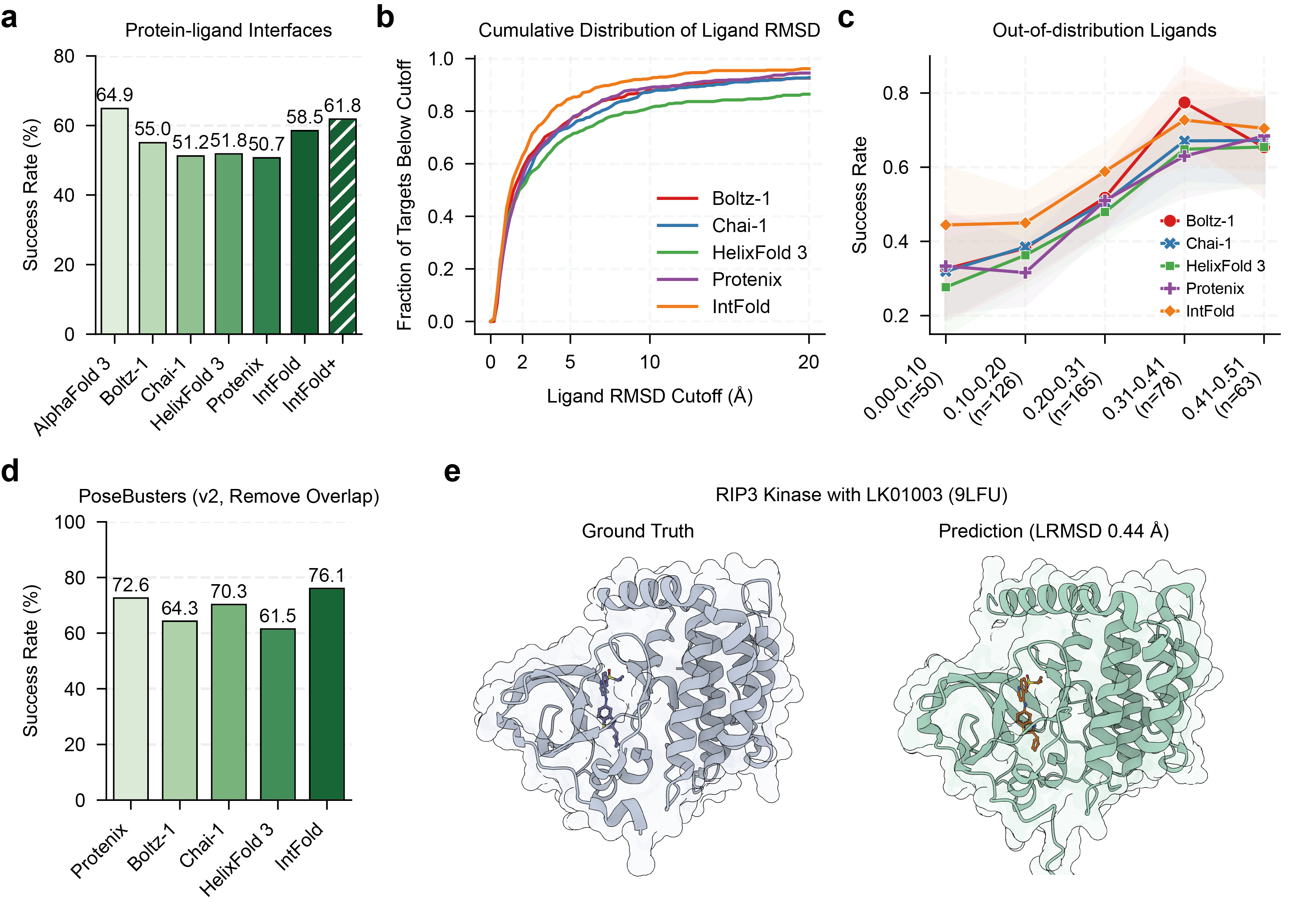} 
\caption{(a) Success rates for protein-ligand interface prediction on FoldBench. (b) Cumulative distribution of ligand RMSD scores, showing the fraction of targets predicted below a given RMSD cutoff. (c) Success rate as a function of ligand similarity to the training set, testing generalization to out-of-distribution molecules. (d) Success rate on the PoseBusters v2 benchmark, filtered for targets not seen during training. (e) Example of a protein-ligand prediction for a recently released RIP3 Kinase}
\label{fig:protein_ligand_bench}
\end{figure}
For protein-ligand interactions, IntFold establishes itself as a highly competitive model on the corresponding FoldBench subset. As shown in Figure 3a, it achieves a success rate of 58.5\%, placing it as the leading model after AlphaFold 3 (64.9\%) and substantially outperforming other methods like Boltz-1 (55.0\%). Moreover, IntFold+ improves the success rate to 61.8\%, further narrowing the gap to AlphaFold 3. A more detailed assessment, based on the cumulative distribution of ligand RMSD, confirms IntFold's superior accuracy over these other competing models at every precision level (Figure~\ref{fig:protein_ligand_bench}b). We also assessed performance as a function of ligand similarity to the training set, which shows that IntFold maintains a significantly higher success rate than other methods for novel ligands, indicating strong generalization capabilities (Figure~\ref{fig:protein_ligand_bench}c).

We also benchmarked on the PoseBusters~\citep{buttenschoen2024posebusters} benchmark. To avoid data leakage, we created a test set by selecting targets from PoseBusters v2 that were deposited after the training data cutoff of 2021-09-30. As shown in Figure~\ref{fig:protein_ligand_bench}d, IntFold achieves a success rate of 76.1\%, outperforming the next best method, Protenix (72.6\%). Additionally, Figure~\ref{fig:protein_ligand_bench}e shows a high-quality prediction for receptor-interacting protein kinase 3 (RIPK3) with its novel inhibitor LK01003.

\subsection{Nucleic Acid Systems}

\begin{figure}[H]
\centering
\includegraphics[width=0.95\linewidth]{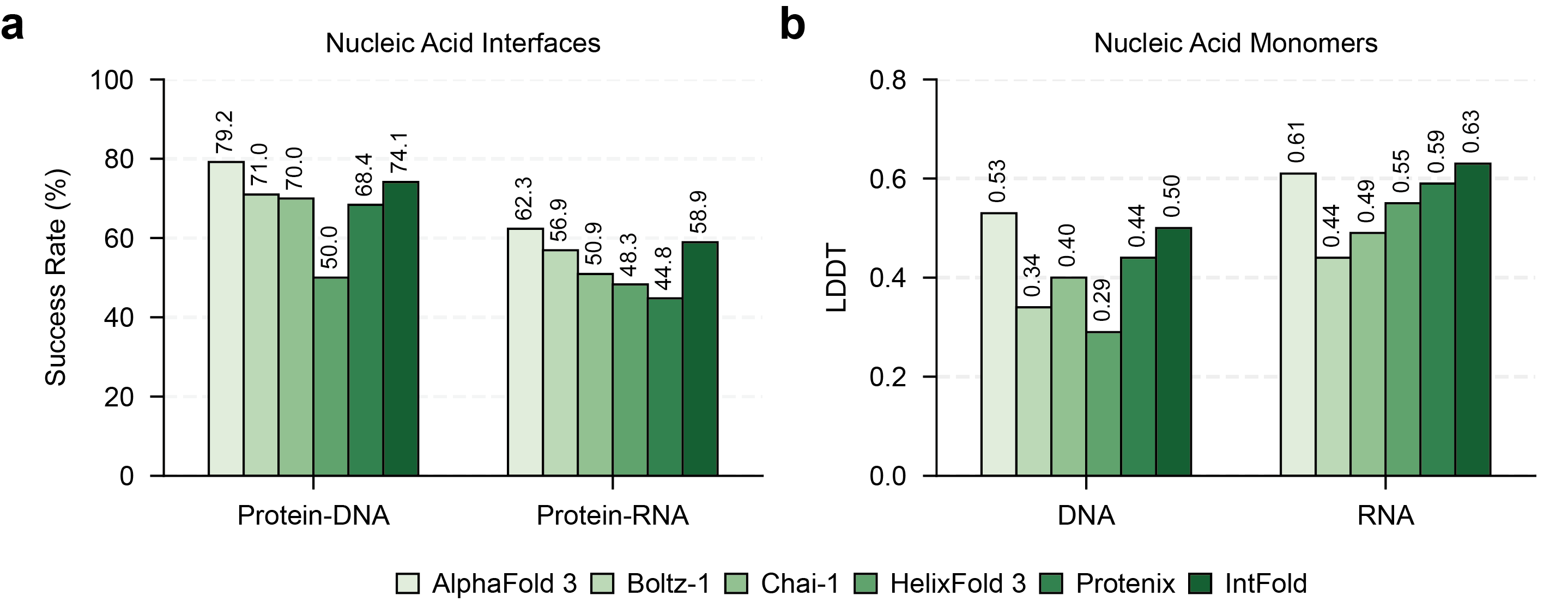} 
\caption{(a) Success rates for protein-nucleic acid interface prediction on FoldBench for both protein-DNA and protein-RNA complexes. (b) Mean LDDT scores for DNA and RNA monomer structure prediction.}
\label{fig:nucleic_bench}
\end{figure}

Finally, we benchmarked IntFold's performance on systems involving nucleic acids, a critical but often challenging modality (Figure~\ref{fig:nucleic_bench}). For protein-DNA interfaces, IntFold's success rate is 74.1\%, outperforming the best contemporary method, Boltz-1 (71.0\%), and demonstrating strong performance compared to AlphaFold 3 (79.18\%). For protein-RNA interfaces, the model achieves a success rate of 58.9\%, again surpassing the next best method, Boltz-1 (56.9\%), and approaching the performance of AlphaFold 3 (62.3\%).

The model also produces high-quality predictions for nucleic acid monomers. For RNA monomers, IntFold's LDDT score of 0.63 surpasses all other models, including AlphaFold 3 with an LDDT score of 0.61. For DNA monomers, IntFold's score of 0.50 is also highly competitive, exceeding other methods like Protenix (0.44) and approaching AlphaFold 3 (0.53). These results confirm IntFold's strength as a general-purpose model with robust capabilities for nucleic acid systems.

\section{Applications}
A key advantage of IntFold is its ability to be efficiently specialized for a range of downstream tasks. This is achieved using the modular adapters described in Section 5.3, which allow for targeted modifications without retraining the entire foundation model. In this section, we demonstrate the power of this approach with three key applications: improving predictions for specific target families, guiding folding with structural constraints, and predicting binding affinity.

\begin{figure}[H]
\centering
\includegraphics[width=\linewidth]{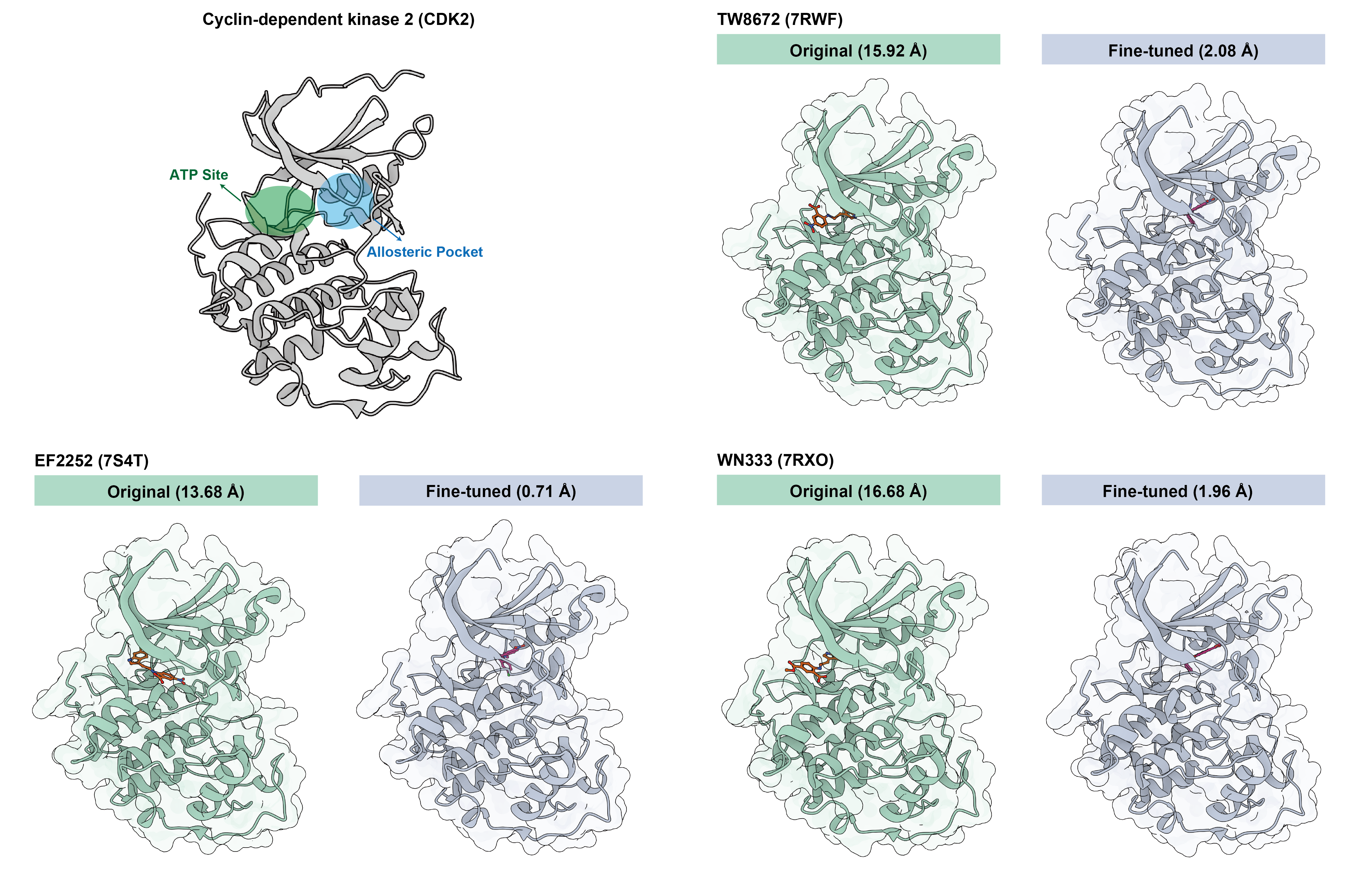} 
\caption{(Top left) Structure of CDK2 showing the ATP site and an allosteric pocket. (Others) Comparison of predictions (LRMSD in parentheses) from the general model (``Original'') and the fine-tuned, specialized model for CDK2 with three different allosteric inhibitors (7RWF, 7S4T, 7RXO).}
\label{fig:cdk2}
\end{figure}

\subsection{Target-type Specific Modeling}

While large models excel at general prediction, achieving high accuracy for specific, therapeutically relevant protein families often requires specialized knowledge. This is especially true for proteins with subtle conformational states that are critical for function. A prime example is the kinase family, particularly proteins like Cyclin-dependent kinase 2 (CDK2), whose activity is modulated by distinct conformations. General-purpose models often fail to capture these inhibitor-induced structural shifts, a significant hurdle for effective structure-based drug design, where modeling these specific states is essential.

To address this challenge, we applied a target-specific adapter to capture these distinct conformations. We used a Low-Rank Adaptation (LoRA) architecture, an efficient method that introduces a small number of trainable parameters while the large base model remains frozen. This adapter was trained on our curated CDK2 dataset, which contains multiple structures of CDK2 bound to a diverse range of inhibitors (Section~\ref{section:data_source}). The results show a clear improvement in predictive accuracy for these specific states. As shown in Figure~\ref{fig:cdk2} and \ref{fig:si_cdk2}, the general model consistently defaults to predicting an open, inactive-like state for CDK2, even in the presence of allosteric inhibitors. In contrast, the fine-tuned model successfully captures the correct closed conformations (7RWF, 7S4T, and 7RXO). This success was quantified across a test set of 40 CDK2 structures. On this set, the general model failed to predict any of the 5 allosteric (closed-state) conformations. The fine-tuned model, however, correctly identified 4 of the 5 allosteric cases, while also maintaining perfect accuracy on the 35 open, inactive-like state structures. This demonstrates that the specialized adapter can robustly identify rare conformational states without compromising its accuracy on more common ones.

\subsection{Guided Folding with Structural Constraints}

\begin{figure}[H]
\centering
\includegraphics[width=\linewidth]{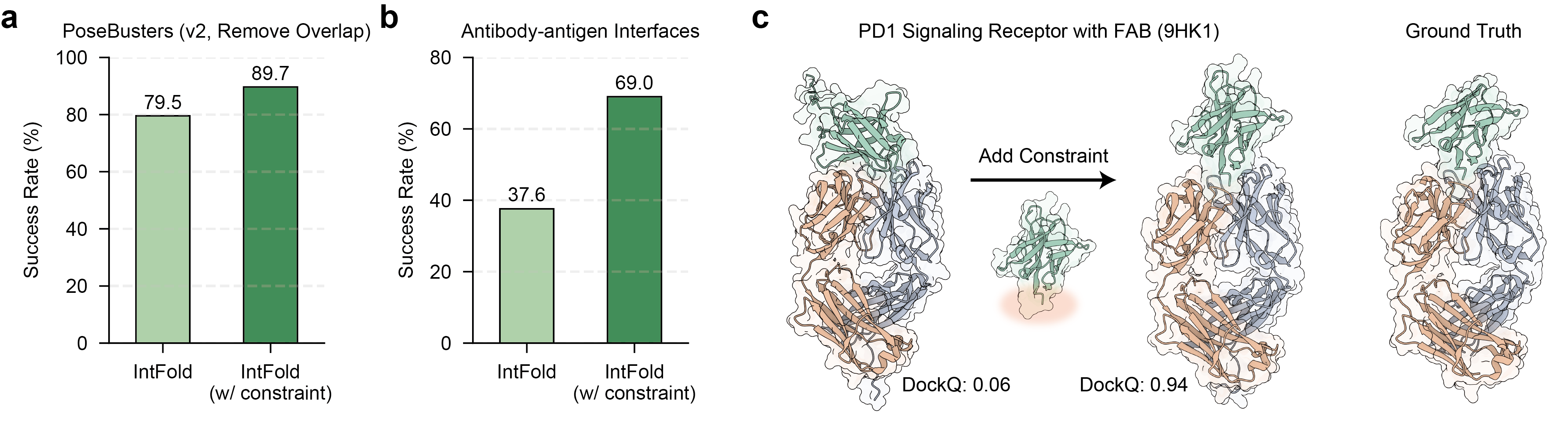} 
\caption{(a, b) Success rate comparison for IntFold with and without structural constraints on the PoseBusters dataset and antibody-antigen interfaces, respectively. (c) A case study on the PD1 signaling receptor with FAB (9HK1), showing how adding a constraint to the binding interface guides the model from an incorrect prediction to the correct docked conformation.}
\label{fig:constraint}
\end{figure}

Structural biologists often have prior knowledge about which residues form an interaction interface, such as a known binding pocket for a ligand or a specific epitope for an antibody. The ability to incorporate this information as a structural constraint is critical not only for improving predictive accuracy but also for experimentally testing specific biological hypotheses about how molecules interact. Providing these constraints allows researchers to guide the model towards a desired configuration, focusing its predictive power on a defined region of interest.

We implemented this capability by training a constraint-specific LoRA adapter. To encode this spatial information for the model, we added a dedicated embedder to the LoRA setup, which processes the known interaction residues. This guided approach leads to dramatic improvements in success rates. For the PoseBusters dataset, applying constraints boosts the success rate from 79.5\% to 89.7\% (Figure~\ref{fig:constraint}a). The effect is even more significant for challenging antibody-antigen interfaces, where the success rate more than doubles from 37.6\% to 69.0\% (Figure~\ref{fig:constraint}b). The case study on the PD1 signaling receptor (9HK1) clearly demonstrates this effect: the unconstrained model produces a physically implausible, incorrect docking pose, but providing the epitope constraint guides the model to the correct docked conformation, perfectly matching the ground truth (Figure~\ref{fig:constraint}c).

\subsection{Predicting Protein-Ligand Binding Affinity}

\begin{figure}[H]
\centering
\includegraphics[width=1\linewidth]{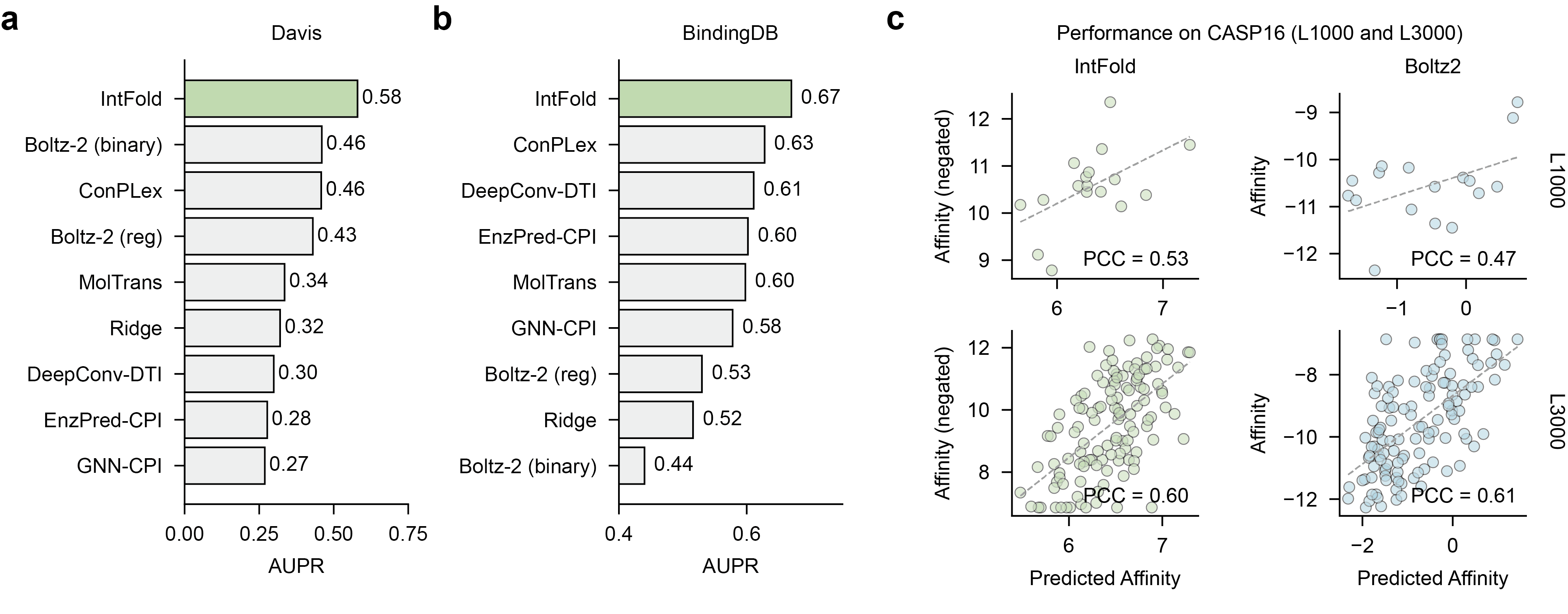} 
\caption{(a, b) Comparison of IntFold against other methods on the Davis and BindingDB benchmarks, as measured by area under the precision-recall curve (AUPR). (c) Correlation between predicted and experimental affinity for IntFold and Boltz-2 on the CASP16 benchmark targets L1000 and L300.}
\label{fig:affinity}
\end{figure}

\begin{table}[h]
\centering
\caption{Performance of IntFold versus Boltz-2 on a subset of the FoldBench benchmark. The test set includes only targets released after January 1, 2024, to avoid data leakage and provide a fair evaluation of generalization performance.}
\label{tab:Boltz-2}
\resizebox{0.8\textwidth}{!}{%
  \begin{tabular}{lccccc c}
    \toprule
    & \multicolumn{5}{c}{\textbf{Interface (Success Rate\,\%)}} & \textbf{Monomer (LDDT)} \\
    \cmidrule(lr){2-6} \cmidrule(l){7-7}
    \textbf{Model} & Prot–Lig & Prot–Prot & Ab–Ag & Prot–DNA & Prot–RNA & Protein \\
    \midrule
    IntFold  & 58.17 & 72.13 & 40.27 & 76.54 & 77.78 & 0.87 \\
    Boltz-2  & 53.90 & 70.54 & 25.00 & 73.84 & 76.92 & 0.88 \\
    \bottomrule
  \end{tabular}%
}
\end{table}

Binding affinity measures how tightly a small molecule attaches to a protein. This measure is critical in drug discovery, as it helps determine whether a potential drug will act on its intended target and be potent enough to produce a therapeutic effect. While structure-based virtual screening is a promising method for discovering drugs against new targets, existing tools lack atomic-level precision and fail to predict binding fitness accurately.

To address the challenge, we trained a post-hoc prediction module on the curated affinity dataset from Section~\ref{section:data_source}. The model's performance was evaluated on the standard DAVIS~\citep{davis2011comprehensive} and BindingDB~\citep{liu2007bindingdb} benchmark datasets, with the area under the precision-recall curve (AUPR) as the primary metric. As presented in Figure~\ref{fig:affinity}a-b, IntFold significantly outperforms a range of existing methods, including both structure-based predictors and various sequence-based approaches.

To further evaluate its performance on novel targets, we tested the model on the recent CASP16 affinity track~\citep{gilson2025assessment}. IntFold's predicted affinities show a stronger correlation with experimental values (Pearson Correlation Coefficient of 0.53) compared to those of Boltz-2 (PCC of 0.47) on L1000 and comparable on L3000 (Figure~\ref{fig:affinity}c). This strong performance extends to a broader head-to-head comparison against Boltz-2~\citep{passaro2025boltz} on FoldBench~\citep{xu2025foldbench} targets released after the training cutoff of Boltz-2 (January 1, 2024). As shown in Table~\ref{tab:Boltz-2}, IntFold consistently outperforms Boltz-2 on most interaction types-for example, reaching a 58.17\% success rate for protein-ligand interactions compared to Boltz-2's 53.90\%-while delivering similar performance for protein monomer folding. Specifically, IntFold shows a protein-ligand success rate of 58.17\% and outperforms Boltz-2's 53.90\%. This consistent strong performance on recent and challenging targets matches the model's reliability for affinity prediction.

\section{Data}
The performance of a foundation model is fundamentally shaped by the scale, diversity, and quality of its training data. Our data strategy was designed to build a comprehensive understanding of biomolecular interactions by integrating multiple sources. We combined high-resolution experimental structures from the Protein Data Bank (PDB) with large-scale, high-quality predicted structures in a process known as distillation. This was supplemented by curating specialized datasets for critical downstream tasks, such as affinity prediction and modeling specific protein families. This multi-faceted approach ensures the model is both a powerful generalist and an adaptable specialist.

\subsection{Data Sources}
\label{section:data_source}
\textbf{Protein Data Bank} For structural data in PDB~\citep{berman2002protein}, we closely followed the processing procedure described in AlphaFold 3~\citep{Abramson2024}. Our weighted PDB dataset is inclusive of all mmCIF files that passed the filtering criteria, including very large or unconventional structures (Figure~\ref{fig:diverse_pdb_examples_appendix}) and structures with identical sequences but distinct docking poses (Figure~\ref{fig:same_sequence_diff_docking_appendix}). Furthermore, we removed all the ions, unknown ligands, and ligands present in AlphaFold 3's exclusion list from the sampling entries.

\textbf{Distillation Datasets} Beyond experimentally resolved structures in PDB, we used three additional distillation datasets.
\begin{itemize}
\item \textbf{Protein Monomer Distillation:} We utilized protein structures from the AlphaFold Database~\citep{varadi2022alphafold} (AFDB). Structures from AFDB were filtered based on 30\% sequence similarity clustering and global pLDDT over 85. Following ESM3~\citep{hayes2025simulating}, we removed structures with fewer than 0.5\(L\) long-range contacts in the chain (where \(L\) is the chain length). This approach was chosen over AlphaFold 3's method of using AlphaFold 2~\citep{jumper2021highly} predictions on MGnify~\citep{mitchell2020mgnify} sequences.

\item \textbf{Disordered Protein PDB Distillation:} We employed the disordered protein PDB distillation set, prepared using AlphaFold-Multimer v2.3~\citep{evans2021protein} predictions on PDB proteins with unresolved regions, consistent with the methodology described in AlphaFold 3. We only used assemblies with fewer than 2000 tokens for this dataset.

\item \textbf{Antibody-Antigen Distillation:} We constructed a dataset of antibody-antigen distillation from PLAbDab~\citep{abanades2024patent}. To ensure diversity, we clustered all sequences at 60\% identity and sampled representative interfaces based on the inverse size of each interface cluster. On the resulting pairs, we performed large-scale structural predictions using IntFold. The distillation set was formed by filtering predictions to examples selected by our ranking method.

\end{itemize}

\textbf{Affinity Dataset} We built a protein–ligand affinity set by merging activity data from ChEMBL~\citep{zdrazil2024chembl}, BindingDB~\citep{liu2007bindingdb}, GalaxyDB~\citep{wang2022structure}, BioLip~\citep{yang2012biolip} and PubChem~\citep{kim2023pubchem}, then cleaning them with the ExCAPE-DB protocol~\citep{sun2017excape}. The pipeline (i) retained compounds whose best IC\textsubscript{50}/K\textsubscript{i}/K\textsubscript{d} value was $\le 10 \mu M$ while keeping matched inactives, (ii) discarded entries flagged as low-confidence or linked to ultra-rare chemotypes, (iii) back-filled missing affinity values from primary sources and filtered out assays that still lacked a number, and (iv) removed protein–ligand pairs whose reported affinities disagreed by more than one log unit across databases. The final dataset comprises roughly three million log-scaled activity measurements spanning 27,000 proteins and 1,000,000 distinct ligands.

\textbf{CDK2 Dataset} For the CDK2 dataset, we selected all entries in the PDB that are explicitly associated with Cyclin-dependent kinase 2. All CDK2-annotated entries in our training set were first partitioned into 10 previously validated allosteric complexes and a larger pool of PDB-derived candidates; after global alignment to a reference orthosteric structure, we calculated the Euclidean distance between each candidate’s dominant pocket centroid and those of the allosteric set, retaining only structures that exceeded our distance and geometry thresholds as orthosteric examples, yielding a final dataset of roughly 200 complexes (\(\sim10\) allosteric + \(\sim190\) orthosteric). The dataset was split at a ligand-similarity threshold of 0.5, and the resulting subset was used to train our model.

\subsection{MSAs and Templates}

\textbf{Genetic Search:} For constructing the paired MSA (providing cross-chain genetic information), we used Jackhmmer~\citep{johnson2010hidden} to search the UniProt~\citep{2023uniprot} database and paired them based on species, following AlphaFold 3. For the main MSA (unpaired sequences), our approach differed from AlphaFold 3. We utilized the ColabFold~\citep{mirdita2022colabfold} search pipeline (typically with MMseqs2~\citep{steinegger2017mmseqs2}) to generate the main MSAs for protein sequences. For RNA sequences, we did not use MSAs.

\textbf{Template Search:} The template search pipeline, including the tools (hmmsearch), databases (PDB sequences), and parameters, was implemented as described in AlphaFold 3.

\section{Modeling}

As illustrated in Figure~\ref{fig:main_figure}, IntFold comprises four main components: an embedding trunk of sequence, a diffusion module, a confidence head, and optional adapters for specialized tasks. The overall model architecture and training procedures are based on the methods described in the AlphaFold 3 supplementary materials~\citep{Abramson2024}. Below, we highlight our novel contributions, including the custom attention kernel, the confidence metric, and insights gained from the model's development.

\subsection{Flash Attention Pair Bias}

\begin{figure}[t]
\centering
\includegraphics[]{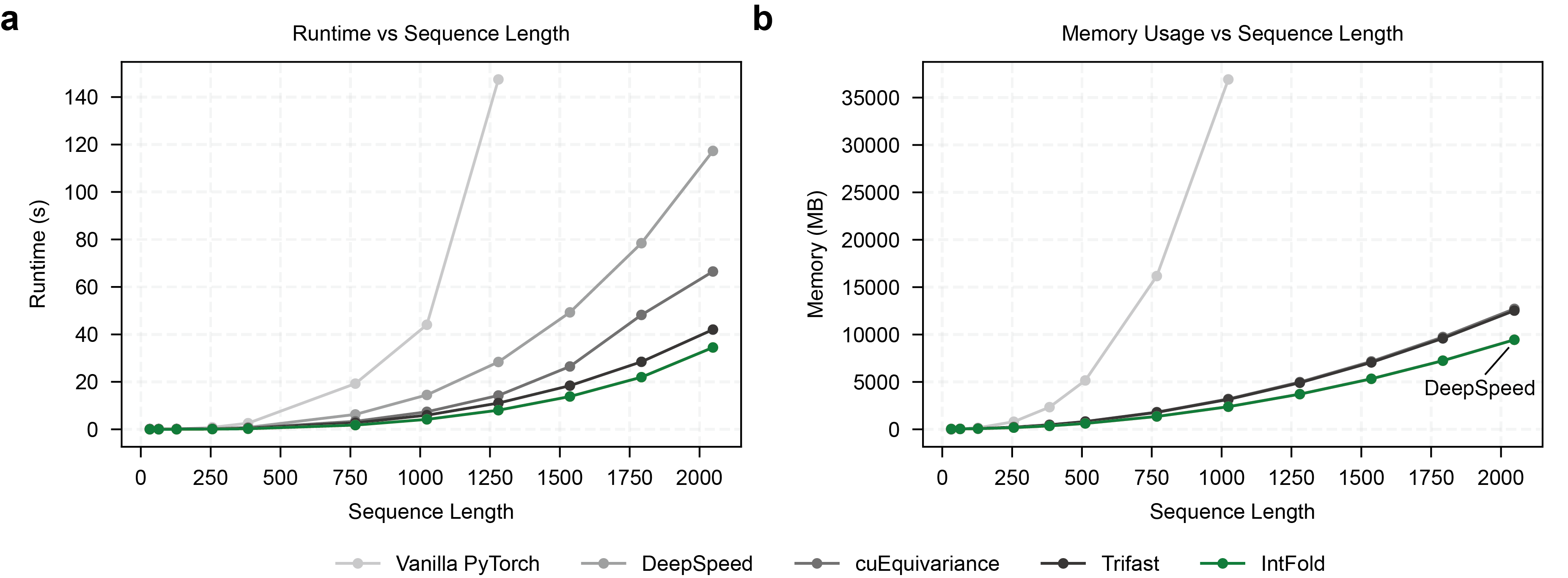} 
\caption{(a) Latency comparison of our custom kernel against other implementations. (b) Memory usage comparison of our custom kernel against other implementations.}
\label{fig:flashpairbiasattention_forward}
\end{figure}

We developed \texttt{FlashAttentionPairBias}, a custom kernel implemented in Triton based on FlashAttention 2~\citep{daoflashattention}. To handle the pair bias, our kernel avoids the memory bottleneck through on-the-fly bias broadcasting. Rather than materializing the full intermediate tensor, only the necessary slice of the bias is loaded from HBM into SRAM during the attention computation. This mechanism brings substantial reductions in peak memory usage and achieves lower forward pass latency than existing implementations, as shown in Figure~\ref{fig:flashpairbiasattention_forward}.

\subsection{Model-Agnostic Ranking}

The stochastic nature of the diffusion process results in variations among predicted structures for a given target. This variability often creates a significant gap between a randomly selected prediction and the best possible structure within a set of generated decoys. While many models employ a learned confidence metric like iPTM to rank predictions, these methods can be tricky to train and are not always reliable for complex interfaces. As a complementary approach, we propose a training-free, model-agnostic ranking method based on structural similarity.

Our method operates on the hypothesis that while incorrect predictions may vary widely, correct predictions tend to be more similar, as they all approximate the ground truth structure. In practice, for a given target, we first generate a diverse set of 25 predictions (5 samples per seed across 5 seeds). We then perform an all-versus-all comparison within this set using DockQ as the pairwise similarity metric. The structure with the highest mean DockQ score against all others is selected as the top-ranked prediction.

This similarity-based ranking method consistently improves performance. When applied to our antibody-antigen benchmark set, this ranking method improves the final success rate by approximately 3\% compared to selecting a random sample. This demonstrates that consensus-based scoring is an effective strategy to better capitalize on the diverse, high-quality structures generated by the model.

\subsection{Modular Adapters}

\label{method:adapters}

We used modular adapters to equip the model with new knowledge and specialized capabilities. These adapters work by introducing a small number of trainable parameters, while the main foundation model remains frozen. We developed two distinct adapter architectures for this purpose.

\textbf{Per-layer LoRA Adapters:} The first architecture uses per-layer Low-Rank Adaptation~\citep{hu2022lora}. This approach is for tasks that steer the model toward specific 3D structures, such as modeling allosteric states. For predictions that require structural constraints like known epitopes, we add an additional embedder to the LoRA setup for encoding this information for the model.

\textbf{Post-Hoc Downstream Module:} The second architecture is a separate post-hoc module designed for downstream tasks different from structure prediction. This external adapter takes the final representations from the model's backbone trunk and processes them through its own parameters. Currently, we use an affinity prediction module composed of four additional Pairformer blocks to output a protein-ligand fitness score.

\subsection{Insights on Instability}

Training a model of this scale and complexity revealed several factors that affect stability, ranging from architectural choices to numerical considerations.

\subsubsection{Activation Explosion and Gradient Spikes}
\begin{figure}[H] 
    \centering

    \includegraphics[width=\textwidth]{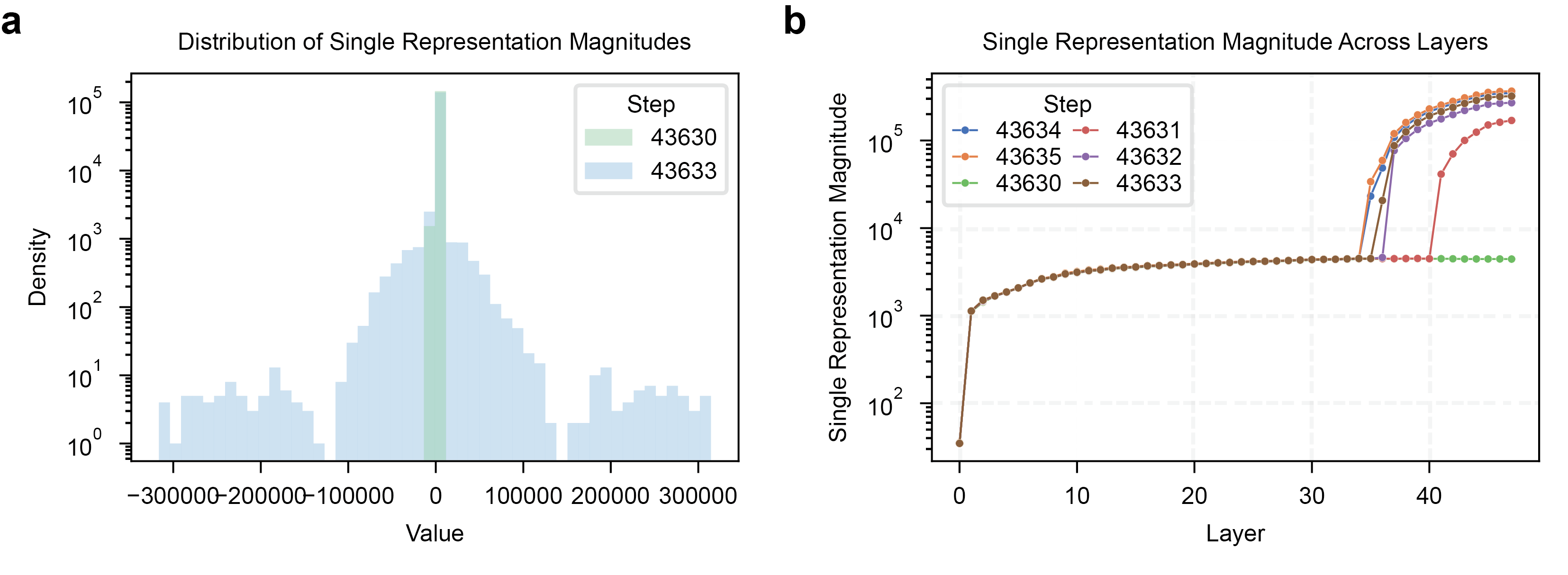}

    \caption{(a) Histogram of single representation magnitudes during training. (b) Single representation magnitude across Pairformer layers at different training steps.}
    \label{fig:single_representation_magnitude_over_layers}
\end{figure}

The primary instability observed was repeated loss spikes, which were preceded by an exploding gradient norm and prevented the model from converging. We identified that these large gradients originated from the backpropagation of the diffusion loss, caused by an abnormally large single representation sent to the diffusion module. This activation explosion initially appears in the Transition module of later layers in the Pairformer Stack, and moves to earlier layers as the training progresses (Figure~\ref{fig:single_representation_magnitude_over_layers}).

To alleviate this, we implemented a ``skip-and-recover'' mechanism. During training, a sample is skipped if the magnitude of its single representation exceeds 40,000. Furthermore, if the gradient norm of a step surpasses a predefined threshold, we will recover training from a previous checkpoint with resampled data batches.

The underlying cause is likely the model's architecture. The pre-Layernorm~\citep{xiong2020layer} design of the Pairformer Stack permits value accumulation on the straight-through path, which is amplified by the Gated Linear Units~\citep{shazeer2020glu}. This hypothesis is supported by experiments where introducing ``sandwich'' LayerNorm~\citep{ding2021cogview} and QK-normalization~\citep{henry2020query} significantly reduced the frequency of loss spikes.

\subsubsection{Parametrization and Initialization}

We also addressed the instability related to parametrization and initialization. For embedding input features like reference atoms, the method described in AlphaFold 3's supplement involves first concatenating features, then applying a single linear projection. We found that this approach can lead to poor single representations, affecting the performance of the model, especially the confidence module. This is because the initialization of a linear layer is related to its input dimension, which becomes very large after concatenation. To solve this, we use separate linear layers for individual features and then add them up. This approach leads to a more stable and reasonable weight initialization, despite being equivalent in forward computation.

Furthermore, we adopted a simplified weight initialization method compared to those described by AlphaFold 2~\citep{jumper2021highly}. We use zero initialization~\citep{goyal2017accurate} for the final layer in residue paths, Kaiming Normal~\citep{he2015delving} for layers with a ReLU activation, and Lecun Normal~\citep{lecun2002efficient} for all others. Additionally, for modules utilizing AdaZero~\citep{peebles2023scalable}, we choose to initialize the output projection of residual blocks with Lecun Normal instead of all zeros.

\subsubsection{Numerical Consideration}

Numerical precision during training is critical as well. While the model's backbone trains effectively and efficiently with Bfloat16 mixed precision, we found that the diffusion module required full Float32 precision for stable training.

We also addressed instability in the loss calculation's alignment step. For the Kabsch alignment, it is better to use the already-augmented ground truth coordinates when aligning with the model's predictions. This is more stable than using the original coordinates, though the two methods are mathematically equivalent. To ensure this alignment is always well-posed, we required the data crops much contain at least four resolved residues, avoiding ambiguous rotational alignment.

\section{Conclusion}

In this report, we introduced IntFold, a versatile and controllable foundation model for biomolecular structure prediction. We have demonstrated that IntFold achieves state-of-the-art predictive accuracy, matching the performance of AlphaFold 3 on critical tasks like protein monomer and protein-protein interaction prediction while significantly outperforming other contemporary methods. This high performance is driven by key technical innovations, including a custom attention kernel that is faster, more memory-efficient than standard industry implementations, and a model-agnostic, training-free, similarity-based ranking method that consistently improves prediction selection.

Beyond its accuracy as a general predictor, IntFold’s primary innovation is its adaptability. Through the use of lightweight, modular adapters, the model can be efficiently specialized for a range of downstream applications critical for drug discovery. We have shown this capability by accurately modeling the inhibitor-specific conformations of CDK2, significantly improving prediction success by incorporating structural constraints, and accurately predicting protein-ligand binding affinity, where it surpasses existing methods on standard benchmarks. 

\section{Limitations and Future Directions}
While IntFold represents a significant step forward, key challenges remain that define our future work.

\textbf{Computational Complexity:} Like other models in its class, IntFold's use of triangle attention presents a computational bottleneck with a complexity of approximately $O(N^3)$, limiting the crop size during training and its speed on very large assemblies. A primary goal is to explore new architectures that can mitigate this complexity without sacrificing accuracy.

\textbf{Accuracy on Challenging Targets:} There is still room to improve predictive accuracy, especially for the most challenging and therapeutically relevant targets like antibody-antigen complexes. While IntFold narrows the performance gap, pushing the boundaries of accuracy for these systems remains a key focus.

\textbf{Expanding Functional Capabilities:} The current work demonstrates several key adaptations. We aim to expand IntFold's capabilities to more downstream applications and into the realm of de novo protein design, further bridging the gap between predicting biomolecular structures and engineering novel functions.

\section{Acknowledgments}
We thank our partner, Jinghe Cloud (Dajing Holding), for providing the stable and scalable computational resources essential for this project. We also thank our colleagues and partners in academia and industry for their invaluable discussions and unwavering support, with special thanks to Shuangjia Zheng and Tao Shen for their insightful comments on the Applications and Modeling sections.

\begingroup
\sloppy 
\printbibliography[heading=bibintoc]
\endgroup
\end{refsection} 

\clearpage
\section*{Appendix}
\appendix

\subsection*{The IntFold Team}

Leon Qiao, Wayne Bai, He Yan, Gary Liu, Nova Xi, Xiang Zhang, and Siqi Sun (project lead).

\subsection*{IntFold+}

To create IntFold+, a model specialized for high-quality antibody-antigen and protein-ligand prediction, we performed a dedicated fine-tuning procedure on the base IntFold model. For this stage, we augmented the training data by including our curated antibody-antigen distillation set~\ref{section:data_source}. We also modified the training protocol by increasing the batch size, removing the MSA pairing information for antibody-antigen complexes to force a greater reliance on structural features, adjusting the ratio of nucleic data, and masking regions with low pLDDT. This combination of data enrichment and a specialized training strategy results in a model highly optimized for the nuances of antibody-antigen docking and protein-ligand binding. While we have demonstrated the effectiveness of these components, a detailed ablation study to quantify the individual contribution of each modification—such as the best data recipe, removal of MSA pairing, and pLDDT masking—remains an area for future work.

\subsection*{Supplementary Figures}

\begin{figure}[H]
    \centering
    \includegraphics[width=0.8\linewidth]{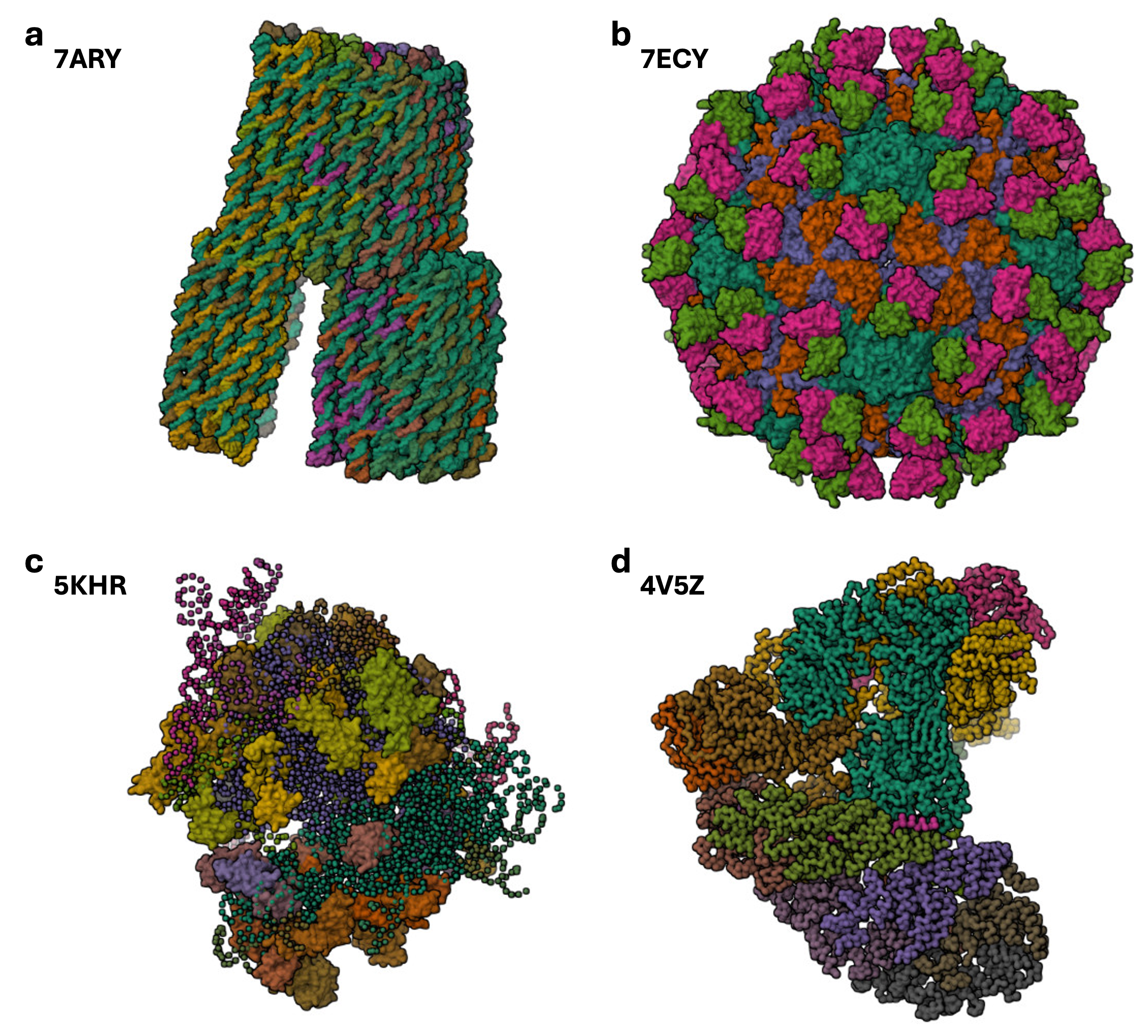}
    \caption{Examples of diverse PDB entries included in the training data. (A) 7ARY, a large DNA-rich assembly. (B) 7ECY, a large virus Icosahedral with FAB. (C) 5KHR, structure with primarily C-alpha resolved atoms. (D) 4V5Z, structure of 80S ribosome with primarily Phosphorus resolved for RNA.}
    \label{fig:diverse_pdb_examples_appendix}
\end{figure}

\begin{figure}
    \centering
    \includegraphics[width=0.8\linewidth]{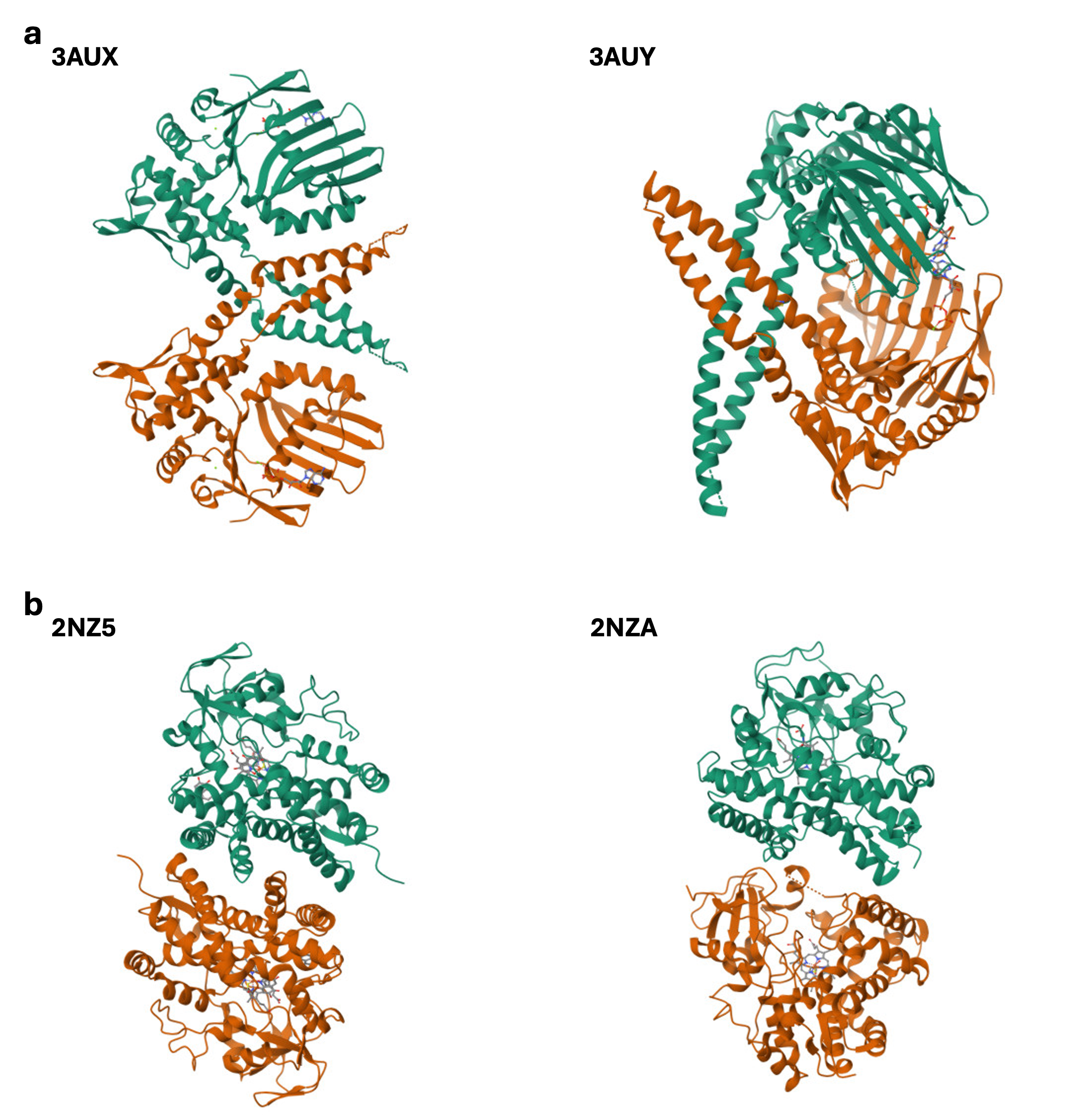}
    \caption{Examples of PDB entry pairs with identical sequences but different experimentally determined docking poses, included in the training data. (a) The pair 3AUX and 3AUY. (b) The pair 2NZ5 and 2NZA.}
    \label{fig:same_sequence_diff_docking_appendix}
\end{figure}

\begin{figure}
    \centering
    \includegraphics[width=0.8\linewidth]{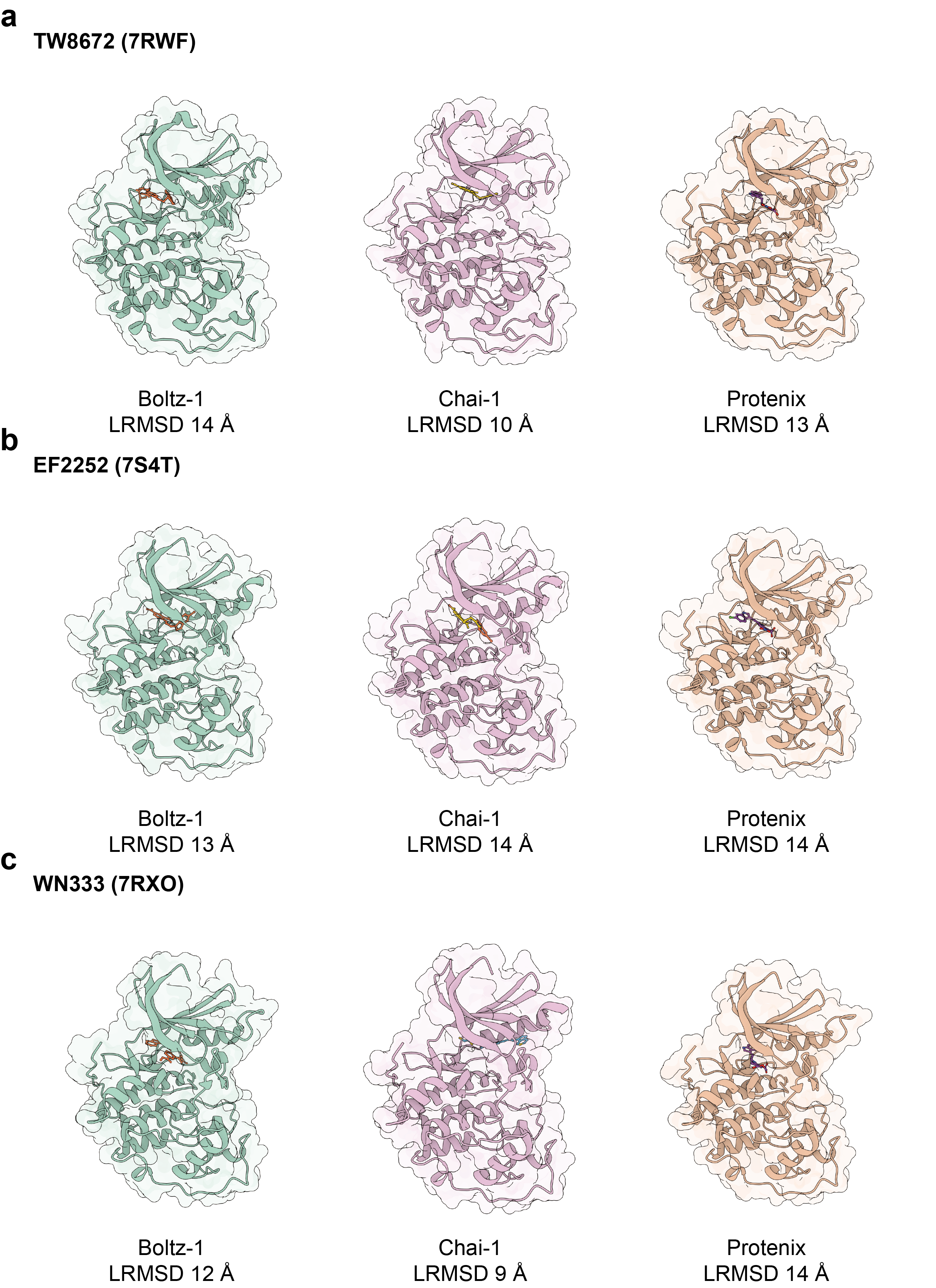}
    \caption{Performance of other methods on CDK2 with three different allosteric inhibitors
(7RWF, 7S4T, 7RXO)}
    \label{fig:si_cdk2}
\end{figure}

\begin{figure}
    \centering
    \includegraphics[width=0.8\linewidth]{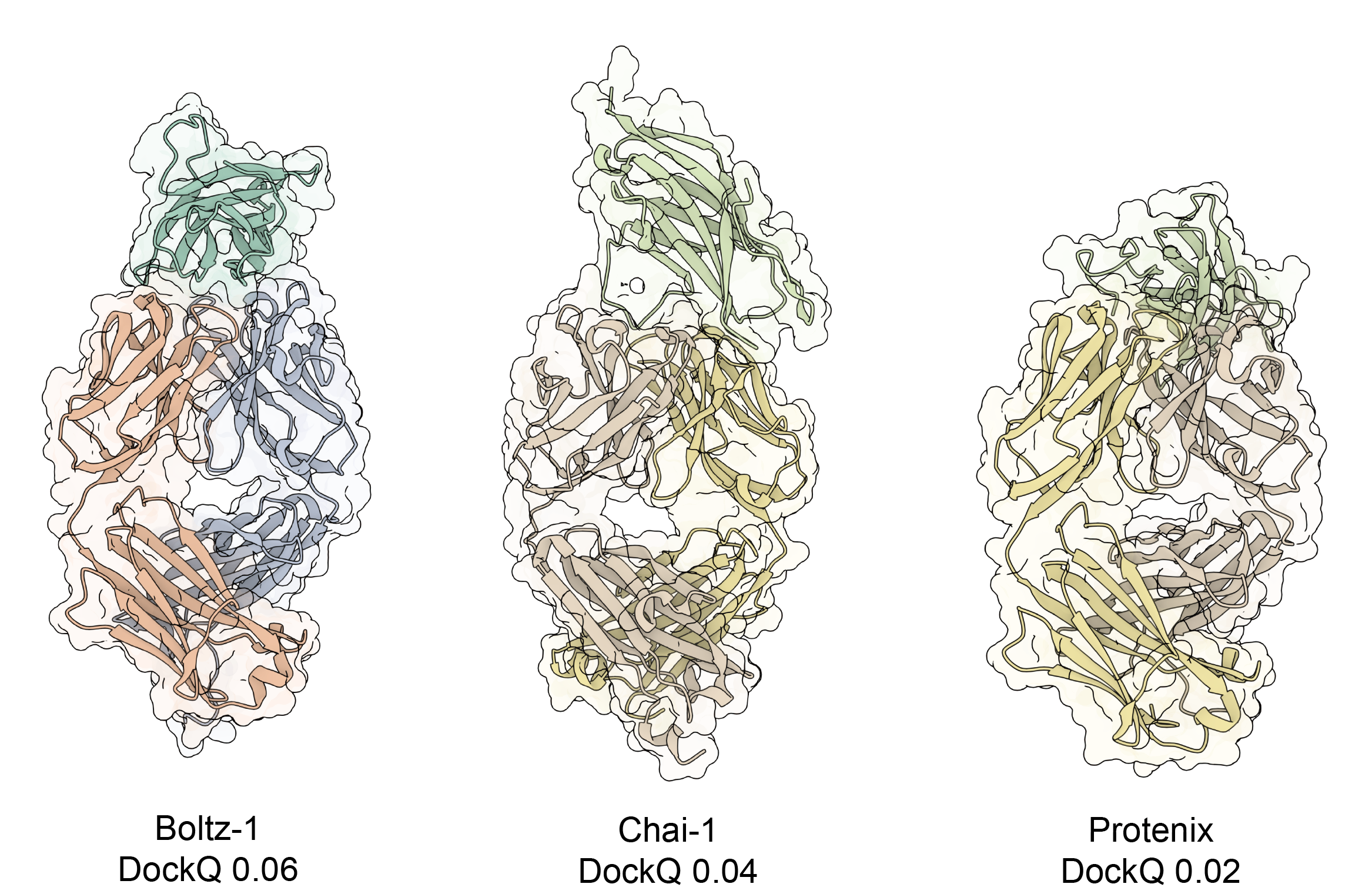}
    \caption{Performance of other methods on a PD1 signaling receptor-FAB complex (9HK1).}
    \label{fig:si_abag}
\end{figure}

\begin{figure}
    \centering
    \includegraphics[width=0.8\linewidth]{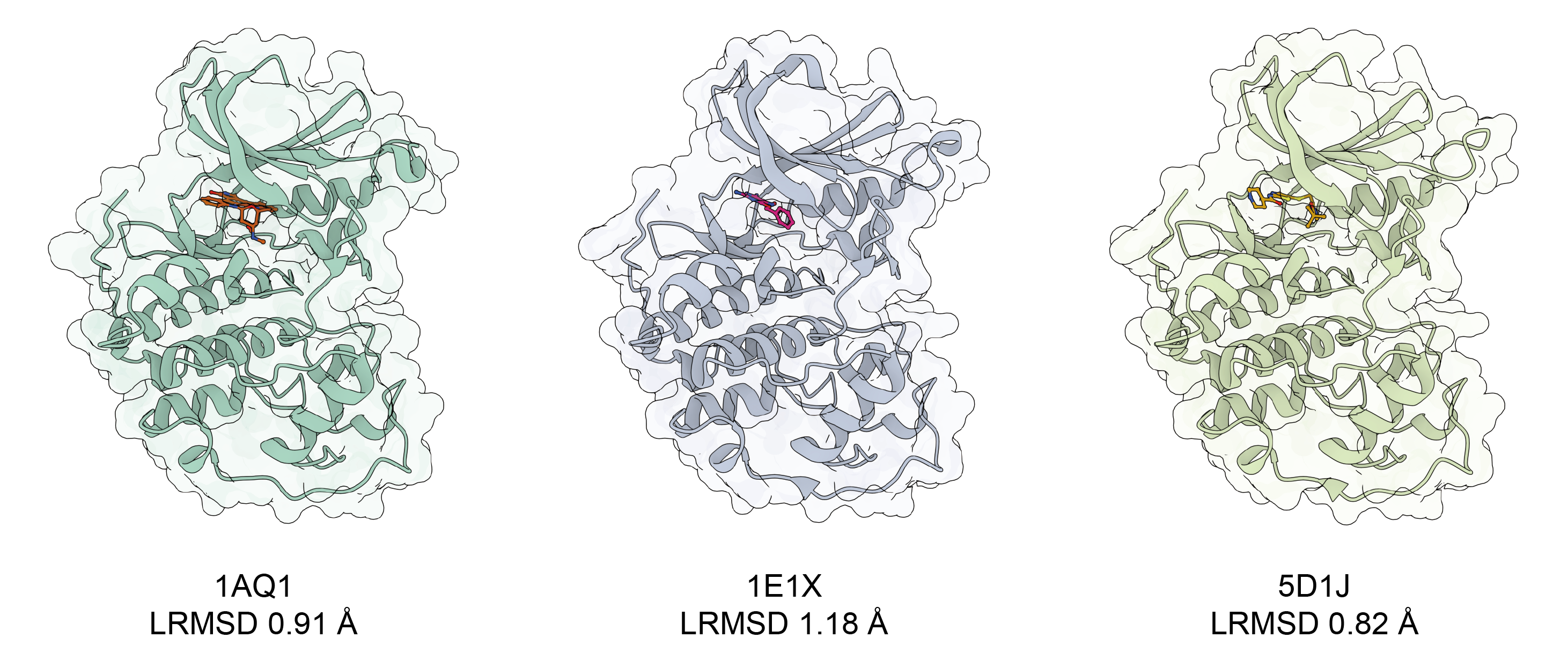}
    \caption{Performance orthosteric sites of CDK2.}
    \label{fig:si_cdk2_orthosteric}
\end{figure}

\begin{refsection}  
\DeclareFieldFormat{labelnumber}{S#1}
\setlength{\labelnumberwidth}{3em}
\end{refsection} 
\end{document}